\documentclass[aps, prb, twocolumn, showkeys]{revtex4-1}

\usepackage{chemformula} 
\usepackage{siunitx} 
\usepackage[inline,shortlabels]{enumitem} 
\usepackage{hyperref} 
\usepackage{graphicx} 

\usepackage{natbib}

\begin{document}

    \title{Selecting the suitable dopants: electronic structures of transition metal and rare earth doped thermoelectric sodium cobaltate}
    \author{M. H. N. Assadi}
    \email{hussein.assadi@unsw.edu.au}
    \email{h.assadi.2008@ieee.org}
    \author{S. Li}
    \author{A. B. Yu}
    \affiliation{School of Materials Science and Engineering, The University of New South Wales, Sydney, NSW 2052, Australia}
    \date{2013}

    \begin{abstract}
        Engineered \ch{Na_{0.75}CoO2} is considered a prime candidate to achieve high-efficiency thermoelectric systems to regenerate electricity from waste heat. In this work, three elements with outmost electronic configurations,
        \begin{enumerate*}[(1)]
        \item an open $d$ shell (Ni),
        \item a closed $d$ shell (Zn), and
        \item a half filled $f$ shell (Eu) with maximum unpaired electrons,
        \end{enumerate*}
        were selected to outline the dopants' effects on electronic and crystallographic structures of \ch{Na_{0.75}CoO2}. Systematic \textit{ab initio} density functional calculations with \ch{DMOL^3} package showed that the Ni and Zn were more stable when substituting Co with formation energy $-2.35$ eV, 2.08 eV when Fermi level equals to the valence band maximum. While Eu is more stable when it substitutes Na having formation energy of $-2.64$ eV. As these results show great harmony with existing experimental data, they provide new insights into the fundamental principle of dopant selection for manipulating the physical properties in the development of high-performance sodium cobaltate based multifunctional materials.
    \end{abstract}
    \maketitle

    \section{Introduction}
        Layered cobaltite \ch{Na_{$x$}CoO2} has recently been subject to intensive research due to its exotic and diverse physical properties. Particularly its unique crystallographic structure is greatly promising for engineering high-efficiency thermoelectric systems.\cite{Fergus2012}
        The triangular nature of \ch{Na_{$x$}CoO2} lattice has an advantage of electronic frustration, creating large spin entropy\cite{Koshibae2001} resulting in large Seebeck coefficient.\cite{Terasaki1997}
        Additionally, in \ch{Na_{$x$}CoO2} electrons propagate in the highly crystalline ionic (\ch{CoO2}) layer\cite{Marianetti2007} while phonons are strongly scattered by the irregular or amorphous arrangement of \ch{Na^{+}} layer.\cite{Foo2004}
        This leads to unprecedented freedom to adjust lattice thermal conduction ($k_i$) and electrical conductivity ($\sigma$) independently in order to achieve a higher figure of merit ($ZT$).\cite{Mahan1996}
        Consequently, \ch{Na_{$x$}CoO2} based thermoelectric materials are among the best candidates for phonon-glass electron-crystal systems.\cite{Snyder2008}
        From a materials engineering viewpoint, controlling Na content has been a prime tool to push the $ZT$ of \ch{Na_{$x$}CoO2} to higher limits.\cite{Wang2004}
        More recently, however, doping has also been utilized to further improve the system's physical properties relevant to $ZT$ to raise the thermoelectric efficiency. For instance, the power factor (PF) of Mg doped \ch{Na_{0.8}CoO2} rose to $\num{0.59e-3}\si{.W.m^{-1}.K^{-2}}$ at $T = 350\si{.K}$ which is $50\%$ higher than the undoped samples.
        Here, Mg dopants increased carriers' mobility by improving the crystallinity of the system at room temperature and thus keeping the resistivity low.\cite{Tsai2012}
        Phonon rattlers such as heavy rare earth elements\cite{Nagira2004} and Ag\cite{Seetawan2006} were also used to decrease $k_i$ as a different strategy to improve $ZT$ of the systems.
        As a result, $3\%$ of Yb dopants significantly increased the PF to $1.5\times 10^{-3}\si{.W.m^{-1}.K^{-2}}$, albeit with a trade-off on increased resistivity.\cite{Nagira2004}
        Furthermore, in order to increase the Seebeck coefficient ($S$), carrier concentration was decreased by doping Ti\cite{Zhang2005}, Ni\cite{Gayathri2006}, Mn\cite{Zhang2006} and Fe.\cite{Dutta2007}
        Among these dopants the best candidate was Ni as $S$ increased with Ni doping from $120\si{.mV/K}$ in the un-doped sample to $150\si{.mV/K}$ in the $2 \si{.at}.\%$ Ni doped sample.\cite{Gayathri2006}
        To continue the progress, fundamental atomistic level understanding of dopant assisted thermoelectric performance in \ch{Na_{$x$}CoO2} should be established.
        Due to the large variation in the bonding nature, electrostatic potential and lattice structures of the \ch{CoO2} and Na layers in \ch{Na_{$x$}CoO2} systems, each of these dopants may be more stable at a particular lattice site.
        This is strongly dominated by the coordination environment, which can be determined using the analysis of x-ray absorption near edge structure.
        However, such a measurement sometimes is limited by the energy range of the instruments.
        Given the complexity and non-trivial interdependency of the electronic transport to the crystallographic structure in the doped \ch{Na_{$x$}CoO2} systems, it is necessary to find the incorporation site of the dopant elements in the particular building block of \ch{Na_{$x$}CoO2}.
        This will establish a roadmap to choose the suitable dopants that reside in a targeted lattice site with specific functionality to separate and control the interdependent parameters important to the enhancement of thermoelectric properties.

        In this work, the behaviour of three different cationic dopants Ni, Zn and Eu in \ch{Na_{0.75}CoO2} host lattice were investigated by \textit{ab initio} calculations.
        These dopants represent an open $d$ shell element, a closed $d$ shell element and an open $f$ shell with maximum magnetic moment element which are typically and frequently used in experiments to manipulate the electric and thermoelectric properties of the \ch{Na_{$x$}CoO2} systems.

    \section{System Settings}
        Spin-polarized all-electrons density functional calculations were performed with \ch{DMol^3} package.\cite{Delley1990, Delley2000}
        Energy calculations were performed with “double-numeric plus polarization” (DNP) basis set.
        Generalized gradient approximation (GGA) based on the Perdew-Wang formalism\cite{Perdew1996} was applied for exchange-correlation functional.
        Real-space global cutoff radii were set for all elements at $5.2\si{.\angstrom}$ to ensure accurate numerical integration for Na orbitals.
        Brillouin zone sampling was carried out by choosing a Monkhorst-Park grid with a spacing of $\sim 0.05\si{.\angstrom^{-1}} $ between $k$ points. Calculated total energy differs only by $10^{-5} \si{.eV/atom}$ when choosing denser grid or larger cutoff radii implying that the results were well converged.
        Finally, the geometry optimization convergence thresholds for energy and Cartesian components of internal forces were set to be $10^{-5} \si{.eV/atom}$ and $0.005 \si{.eV/\angstrom}$ respectively.

        The lattice parameters of \ch{NaCoO2} (with $100\%$ of Na occupancy) were found to be $2.87\si{\angstrom}$ for $a$ and $10.90\si{\angstrom}$ for $c$ which are in reasonable agreement with experimental lattice parameters\cite{Chen2004} as the difference is less than $0.2\%$.
        First, the Na ions patterning in the \ch{Na_{0.75}CoO2} was investigated.
        A $4a\times2a\times1c$ supercell of \ch{NaCoO2} was constructed.
        Then 4 Na vacancies out of 16 total Na sites of the \ch{NaCoO2} supercell were created by removing 2 Na ions from the upper Na layer, Z = 0.75, and 2 Na ions from the lower Na layer, Z = 0.25, resulting in $75\%$ Na occupancy (Z is the fractional coordinate along $c$ direction).
        While fixing the lattice constants to the theoretical values, the internal coordinates of all ions in the supercell were relaxed to find the final equilibrium structure for \ch{Na_{0.75}CoO2}.
        Out of numerous possibilities of the Na ions distribution patterns, 80 structures were considered for calculations to find the ground state configuration.
        In the fully relaxed \ch{Na_{0.75}CoO2} structure the ratio of Na ions occupying Na2 site to the ones occupying Na1 (Na2/Na1) sites was 5 which is in agreement with the previous DFT calculations.\cite{Meng2008}
        Additionally, the shortest \ch{Na}--\ch{Na} separation in each of the Na layers that were located at Z = 0.25 and Z = 0.75 was $2.84\si{.\angstrom}$.

\begin{figure*}[!tp]
        \begin{minipage}{\textwidth}
            \centering
            \includegraphics[width=0.9\textwidth]{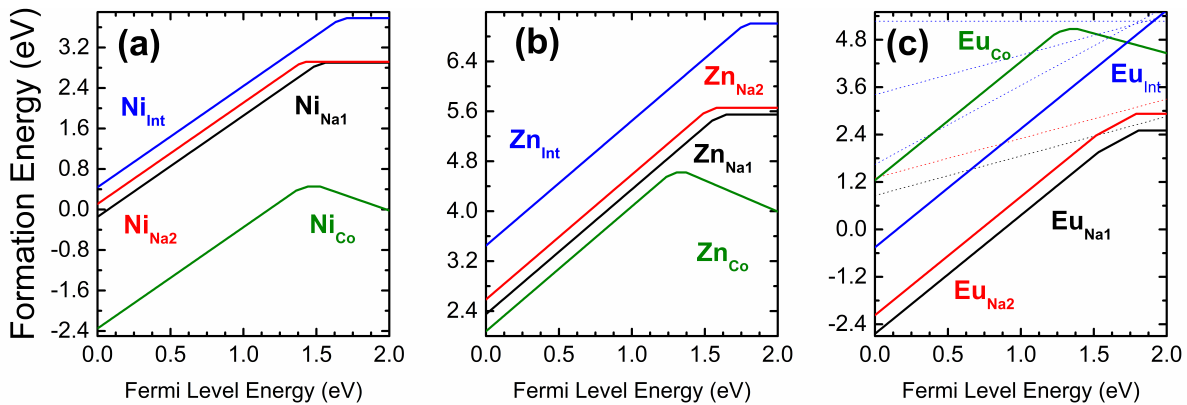}
            \caption{\label{fig:1}The Formation energy ($E^f$) of Ni, Zn, and Eu dopants in \ch{Na_{0.75}CoO2} is given in part (a), (b) and (c) respectively. $E^f$ is presented as a function of Fermi level. The dashed lines present negative U charge states that are not stable for any range of $E_{\mbox{\scriptsize{Fermi}}}$.}
        \end{minipage}
        \end{figure*}

        Ni and Zn have an oxidation state of 2+ while Eu can be stable as either \ch{Eu^{2+}} or \ch{Eu^{3+}}.
        This implies that these dopants may have a diverse electrical activity depending on both incorporation site in the host lattice of \ch{Na_{0.75}CoO2} and on their charge state.
        Since only cationic dopants are being considered here, dopants' formation energy ($E^f$) was calculated for four possible geometric configurations:
        \begin{enumerate*}[(a)]
        \item when the cationic dopant substitutes a Na ion at Na1 site,
        \item when a cationic dopant substitutes a Na ion at Na2 site,
        \item when a cationic dopant occupies an interstitial site in Na layer and
        \item when it substitutes a Co ion.
        \end{enumerate*}
        Since the ionic volume of the considered dopants was larger than the interstitial cavity in the \ch{CoO2} layer, this interstitial site was not considered.
        The formation energy of dopants ($E^f$) as a function of the Fermi level energy ($E_{\mbox{\scriptsize{Fermi}}}$) was calculated according the standard procedure.\cite{vandeWalle2004}
        The chemical potentials of dopant X, Co and Na were calculated as $\mu = \left[E^t \left(\ch{X_nO_m}\right) - \tfrac{1}{2}m E^t \left(\ch{O2}\right)\right]/n$, where $E^t$ is DFT's total energy and \ch{X_nO_m} is the most stable oxide of the respective elements.
        The $E^f$ of all investigated elements is presented in Fig. \ref{fig:1} (a)--(c) for comparison.

    \section{Results}
        \subsection{\ch{NaCoO2:Ni}}
\begin{figure}
                \centering
                \includegraphics[width=0.9\columnwidth]{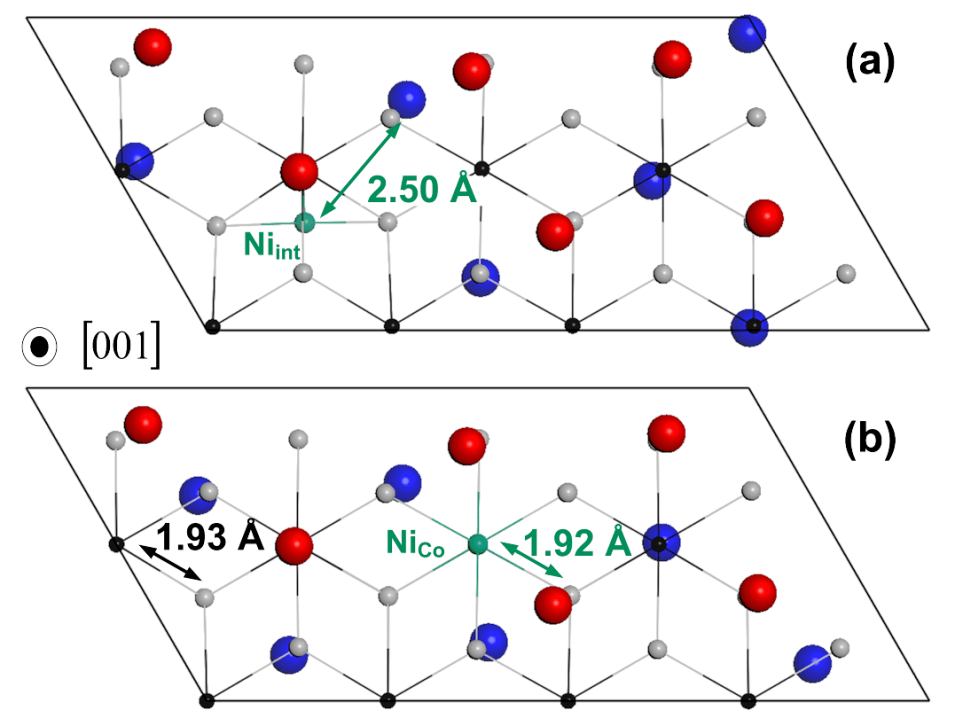}
                \caption{\label{fig:2}Schematic representation of Ni doped \ch{Na_{0.75}CoO2}. Parts (a) and (b) demonstrate the ionic arrangement of \ch{Na_{0.75}CoO2} when doped with \ch{Ni_{Int}^{$\cdot\cdot$}} and \ch{Ni_{Co}^{$\cdot\cdot$}} respectively. \ch{Ni_{Co}^{$\cdot\cdot$}}  and  \ch{Ni_{Int}^{$\cdot\cdot$}} are the most and least stable Ni configurations when $E_{\mbox{\scriptsize{Fermi}}}$ = VBM as the case of p-type \ch{Na_{0.75}CoO2}.}
            \end{figure}

            The formation energy of Ni dopants is presented in Fig. \ref{fig:1} (a).
            When $E_{\mbox{\scriptsize{Fermi}}}$ is located at the valence band maximum (VBM), the formation energy of \ch{Ni_{Co}^{$\cdot\cdot$}} is the lowest among all configurations with a value of $-2.35$ while \ch{Ni_{Na 1}^{$\cdot\cdot$}}, \ch{Ni_{Na 2}^{$\cdot\cdot$}} and \ch{Ni_{Int}^{$\cdot\cdot$}} have relatively higher $E^f$ of $-0.15$ eV, 0.11 and 0.44 eV.
            As $E_{\mbox{\scriptsize{Fermi}}}$ moves across the band gap toward higher values, the order of $E^f$ of different configuration remains the same although the difference between the $E^f$ of \ch{Ni_{Na 1}^{$\times$}} and \ch{Ni_{Na 2}^{$\times$}}  vanishes.
            Additionally, all Ni configurations remain in their highest charge state for wide range of $E_{\mbox{\scriptsize{Fermi}}}$.
            \ch{Ni_{Co}}'s (2+/1+) transition occurs at $E_{\mbox{\scriptsize{Fermi}}}$ = 1.36 eV. \ch{Ni_{Na 1}} and \ch{Ni_{Na 2}} have their (2+/1+) transitions at 1.48 eV and 1.38 eV and \ch{Na_{Int}}'s $\epsilon$(2+/1+) transition is at 1.63 eV.
            When $E_{\mbox{\scriptsize{Fermi}}}$ = 2 eV, all Ni configurations are at their lowest charge state i.e.\ neutral for Ni dopant located at Na layer and single negative for Ni substituting Co (\ch{Ni$'$_{Co}}).
            For the entire range of considered $E_{\mbox{\scriptsize{Fermi}}}$, \ch{Ni_{Co}} is more stable by at least 2 eV than all other configurations.
            Since \ch{Ni_{Int}} and \ch{Ni_{Co}} are the least and most stable configuration, their geometric and electronic structure are examined here.
            A schematic representation of \ch{Ni_{Int}} and \ch{Ni_{Co}} lattice configurations is presented in Fig. \ref{fig:2}(a) and (b) respectively.
            In the case of \ch{Ni_{Int}} as in Fig. \ref{fig:2}(a), the distance of Ni from its nearest Na neighbor is $2.50\si{.\angstrom}$.
            This distance is substantially larger than the sum of ionic radii of \ch{Ni^{2+}} ($0.69 \si{.\angstrom}$) and  \ch{Na^{+}} ($1.02\si{.\angstrom}$), indicating a large Coulombic repulsive force acting upon the ions in the sodium layer with no chemical bond being formed.
            However, this distance is shorter than the shortest \ch{Na}--\ch{Na} distance in pristine \ch{Na_{0.75}CoO2} of $2.84\si{.\angstrom}$, indicating ionic re-arrangement in Na layer to accommodate for \ch{Ni_{Int}}.
            One can also notice that in the pristine \ch{Na_{0.75}CoO2} the ratio of Na2/Na1 was 5, however, for \ch{Na_{0.75}CoO2:Ni_{Int}}, this ratio is 1.17.
            The reduction in the concentration of Na2 is caused by the creeping of Na ions from the vicinity of \ch{Ni_{Int}} to the next local minima of the electrostatic potential at Na1 sites.
            As shown in Fig. \ref{fig:2}(b), \ch{Ni_{Co}} does not cause any significant ionic re-arrangement in comparison to \ch{Ni_{Int}}.
            The only notable change is a minor shrinkage in \ch{Ni_{Co} - O} bond ($1.92\si{.\angstrom}$) in comparison to the \ch{Co - O} bond ($1.93\si{.\angstrom}$) which is caused by \ch{Ni^{2+}}'s smaller radius of $0.69\si{.\angstrom}$ than the radius of \ch{Co^{3+}}, $0.75\si{.\angstrom}$, at its low spin configuration.
            It is evident that \ch{Ni_{Co}} caused less lattice distortion than \ch{Ni_{Int}} when compared with the pristine structure.
            This may be one of the stabilizing causes of this configuration.

\begin{figure}
                \centering
                \includegraphics[width=0.9\columnwidth]{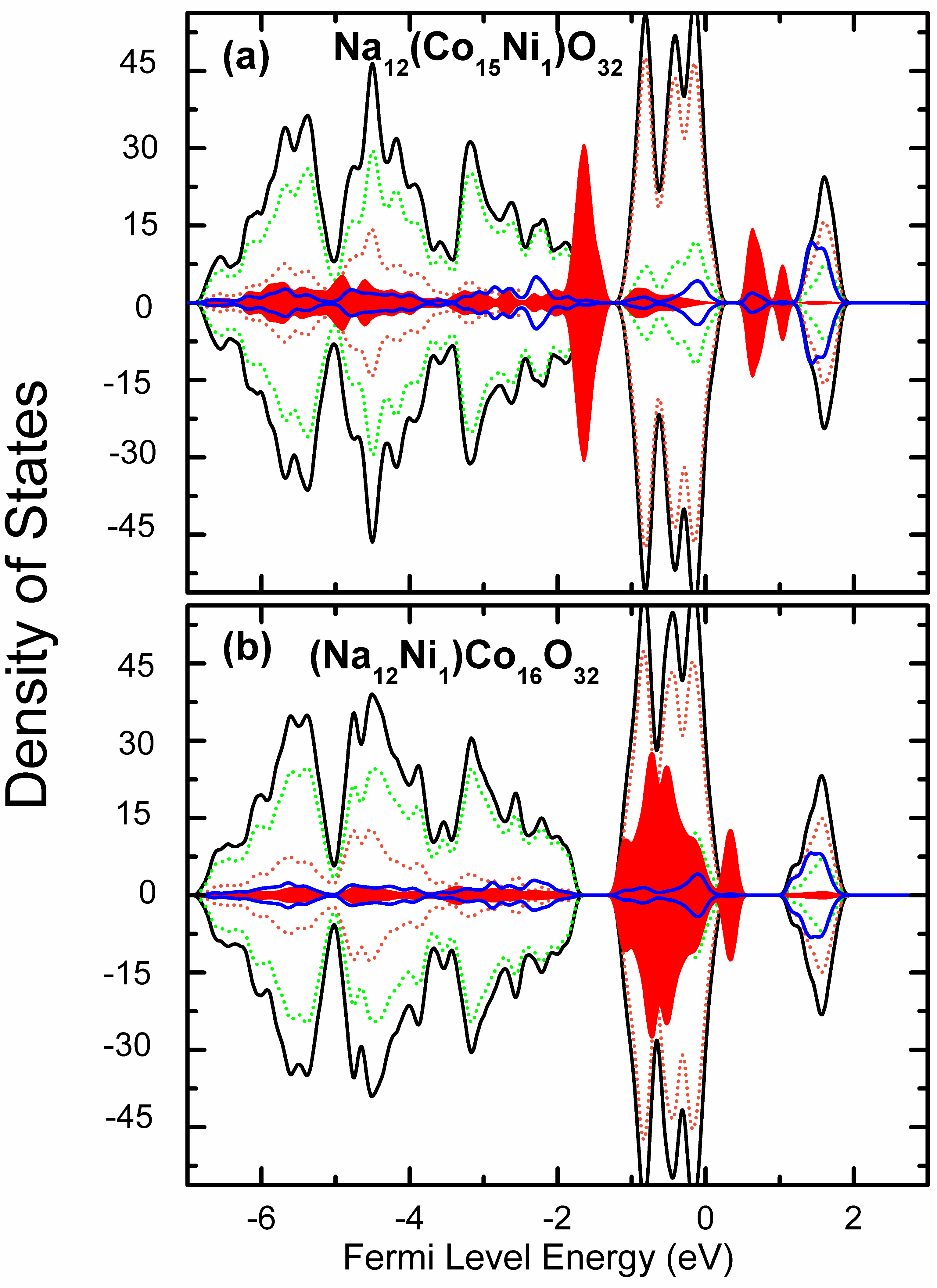}
                \caption{\label{fig:3}Total and partial density of states (PDOS) of \ch{Ni_{Co}^{$\cdot\cdot$}}  and  \ch{Ni_{Int}^{$\cdot\cdot$}} doped \ch{Na_{0.75}CoO2} presented in part (a) and (b) respectively. Black, green, orange and blue lines correspond to the total, O 2$p$, Co $3d$ and Na $3s$ respectively. The red shaded area represents Ni $3d$ states. For the purpose of clarity, the PDOS of Ni and Na have been magnified.}
            \end{figure}

            Partial densities of states (P/DOP) of Ni $3d$ electrons along with the P/DOS of host lattice ions, i.e.\ Co $3d$, O $2p$ and Na are presented in Fig. \ref{fig:3}.
            As shown in Fig. \ref{fig:3}(a), \ch{Ni_{Co}} $3d$ states are mainly localized in two different regions of the band structure.
            The \ch{Ni_{Co}} $3d$ states also stretch after 2 eV below the valence band maximum.
            These peaks represent the bonding states of \ch{Ni_{Co}} and indicate strong hybridization between O and Ni which is partially responsible for the relative stability of \ch{Ni_{Co}} in \ch{Na_{0.75}CoO2}. Then, there are two major peaks in the conduction band that represent the antibonding states. Antibonding states usually remain empty in the absence of other electrically active dopants. \ch{Ni_{Int}} $3d$ states [Fig. \ref{fig:3}(b)] distribute differently in the band structure. These states are spread through the valence band and share the same characteristic of Na $3s$ states, behaving as electron donors. Since the VBM is located overly close to the CBM, it is predicted that Ni dopants will adapt their highest charge state of 2+ for the entire range of permissible $E_{\mbox{\scriptsize{Fermi}}}$.

        \subsection{\ch{NaCoO2:Zn}}
\begin{figure}
                \centering
                \includegraphics[width=0.9\columnwidth]{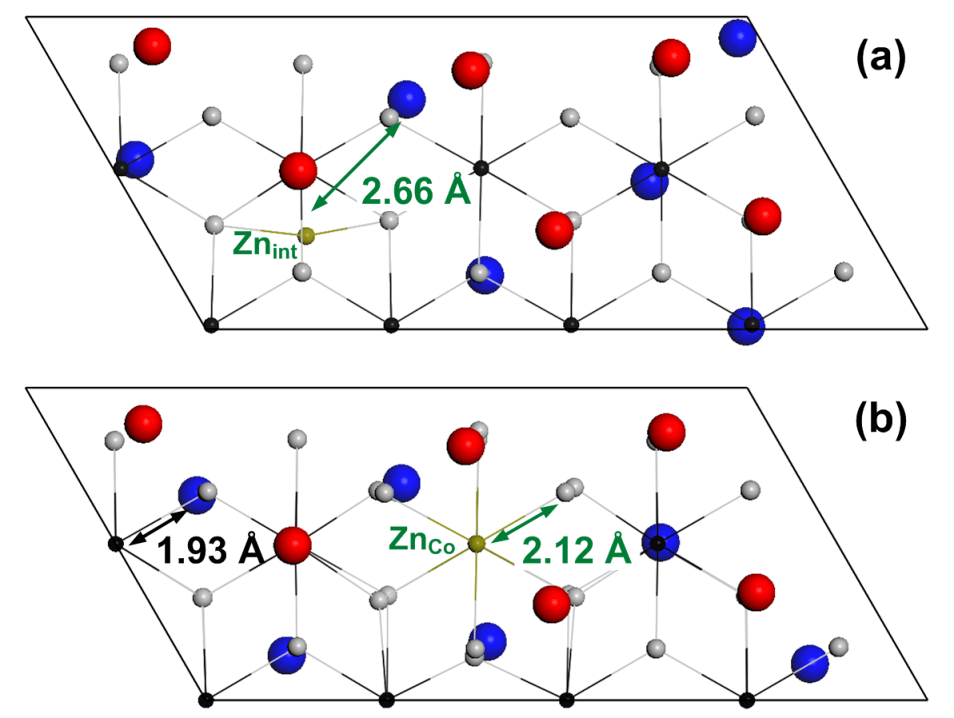}
                \caption{\label{fig:4} Schematic representation of Zn doped \ch{Na_{0.75}CoO2}. Parts (a) and (b) demonstrate the ionic arrangement of \ch{Na_{0.75}CoO2} when doped with  \ch{Zn_{Int}^{$\cdot\cdot$}} and \ch{Zn_{Co}^{$\cdot\cdot$}} respectively.  \ch{Zn_{Co}^{$\cdot\cdot$}} and \ch{Zn_{Int}^{$\cdot\cdot$}} are the most and least stable Zn configurations when $E_{\mbox{\scriptsize{Fermi}}}$ = VBM as the case of p-type \ch{Na_{0.75}CoO2}.}
            \end{figure}

            The formation energy of Zn dopants is presented in Fig. \ref{fig:1}(b).
            When $E_{\mbox{\scriptsize{Fermi}}}$ = VBM, the $E^f$ of \ch{Zn_{Co}^{$\cdot\cdot$}}, the lowest of all configurations, is 2.08 eV while \ch{Zn_{Na 1}^{$\cdot\cdot$}}, \ch{Zn_{Na 2}^{$\cdot\cdot$}} and \ch{Zn_{Int}^{$\cdot\cdot$}} have successively higher $E^f$ of 2.35 eV, 2.59 eV and 3.44 eV respectively.
            As $E_{\mbox{\scriptsize{Fermi}}}$ moves across the band gap toward the higher energies at conduction band minimum, the $E^f$ of \ch{Zn_{Co}^{$\cdot\cdot$}} remains the lowest among all configurations.
            For values of $E_{\mbox{\scriptsize{Fermi}}}$ in the vicinity of $\sim 1.23\si{.eV}$, \ch{Zn_{Co}} undergoes frequent charge transformations to lower charge states and for $E_{\mbox{\scriptsize{Fermi}}}$ = 1.37 eV, \ch{Zn_{Co}} takes negative charge state.
            As the least and most stable configurations, a schematic representation of \ch{Na_{0.75}CoO2:Zn_{Int}} and \ch{Na_{0.75}CoO2:Zn_{Co}} crystal geometry is presented in Fig. \ref{fig:4}(a) and (b) respectively.
            In the case of \ch{Zn_{Int}} as shown in Fig. \ref{fig:4}(a), the distance of Zn from its nearest Na neighbor is $2.66\si{.\angstrom}$.
            Similar to the \ch{Na_{0.75}CoO2:Ni_{Int}} system, this distance is substantially larger than the sum of ionic radii of \ch{Zn^{2+}} ($0.74\si{.\angstrom}$) and \ch{Na^{+}} ($1.02\si{.\angstrom}$), signifying the role of large Coulombic repulsive force ionic placement in the crystal.
            Again, this distance is shorter than the closest \ch{Na}--\ch{Na} distance in pristine \ch{Na_{0.75}CoO2} of $2.84\si{.\angstrom}$ which demonstrates the ionic re-arrangement in Na layer to accommodate for \ch{Zn_{Int}} as evidenced by the ratio of Na2/Na1 in \ch{Na_{0.75}CoO2:Zn_{Int}}, that is 1.17.
            Fig. \ref{fig:4}(b) demonstrates that \ch{Ni_{Co}} causes expansion of $9.84\%$ in \ch{Zn_{Co} - O} bond ($2.12\si{.\angstrom}$) when compared to the original \ch{Co - O} bond ($1.93\si{.\angstrom}$) which is caused by Zn's larger radius of $0.90\si{.\angstrom}$ than the radius of \ch{Co^{3+}}, $0.75\si{.\angstrom}$, at its low spin configuration.
            However, such a difference of ionic radii between Zn and Co does not promote the rearrangement of Na ions as the Na2/Na1 ratio remains intact in \ch{Na_{0.75}CoO2:Zn_{Co}}.
            Once again, it is evident that \ch{Zn_{Co}} caused less lattice distortion than \ch{Zn_{Int}} when compared with the pristine structure resulting in higher stabilization.

            \begin{figure}
                \centering
                \includegraphics[width=0.9\columnwidth]{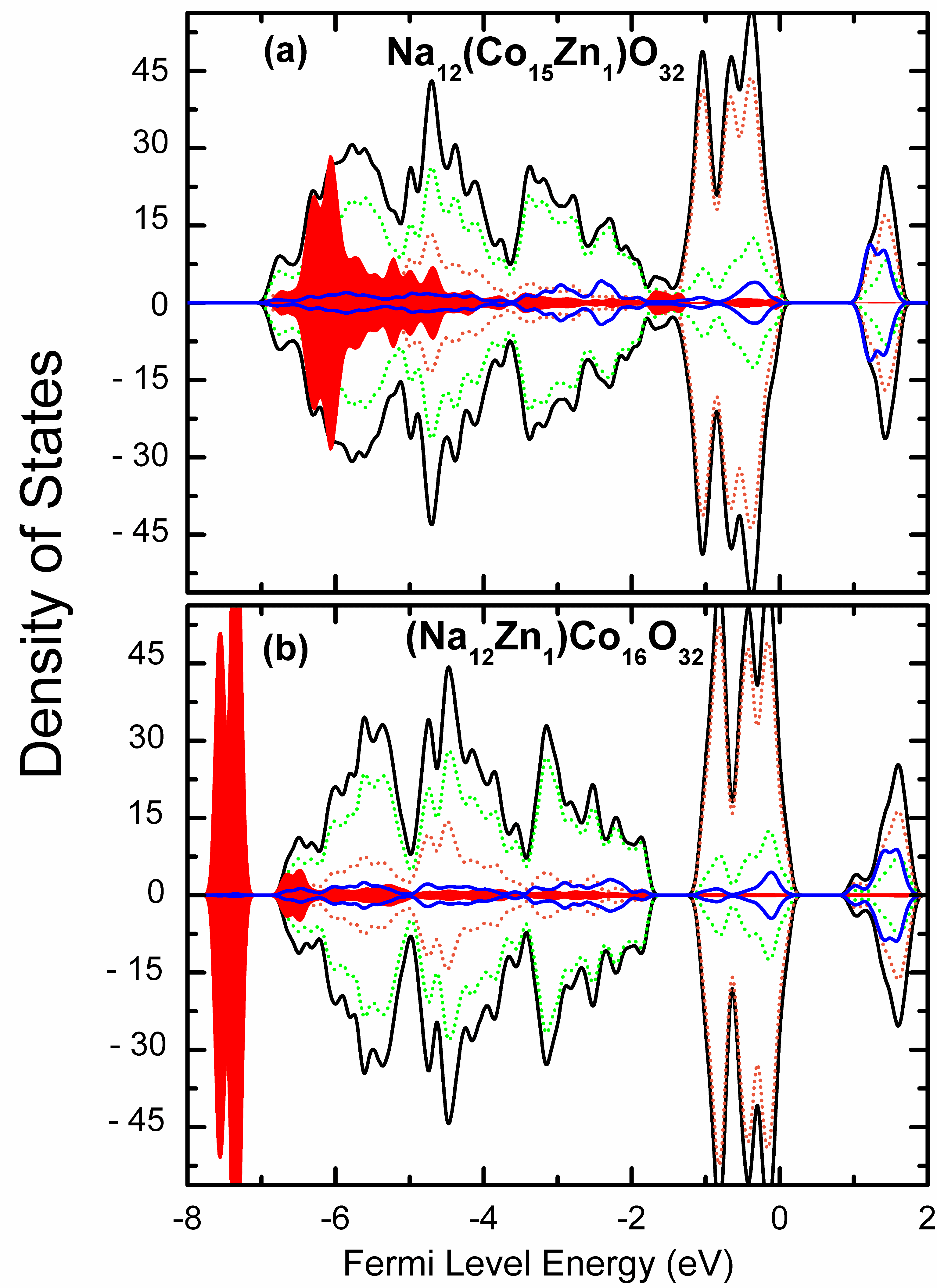}
                \caption{\label{fig:5}Total and partial density of states (PDOS) of \ch{Zn_{Co}} and \ch{Zn_{Int}} doped \ch{Na_{0.75}CoO2} presented in part (a) and (b) respectively. Black, green, orange and blue lines correspond to the total, O $2p$, Co $3d$ and Na $3s$ respectively. The red shaded area represents Zn $3d$ states. For the purpose of clarity, the PDOS of Zn and Na have been magnified.}
            \end{figure}

            Partial densities of states (P/DOP) of Zn $3d$, Co $3d$, O $2p$ and Na $3s$  are presented in Fig. \ref{fig:5}.
            As shown in Fig. \ref{fig:5}(a), \ch{Zn_{Co}} $3d$ states are mainly spread at the bottom of the valence band over the range of $\sim 4$ to $\sim 7\si{.eV}$ below the Fermi level.
            These are \ch{Zn_{Co}} bonding states and show strong hybridization with O and Co states.
            The magnitude of the hybridization shows the strength of the crystal field acting on the Zn ions which is imposed by O ligands.
            \ch{Zn_{Int}} $3d$ states, as shown in Fig. \ref{fig:5}(b) are sharply localized in a separate impurity band which is located $\sim 1\si{.eV}$ below the bottom of the valence band.
            Such an electronic distribution indicates that unlike the case of Ni with partially filled $d$ shell, Zn $3d$ states do not participate in the conduction mechanism.

        \subsection{\ch{Na_{0.75}CoO2:Eu}}
            \begin{figure}
                \centering
                \includegraphics[width=0.9\columnwidth]{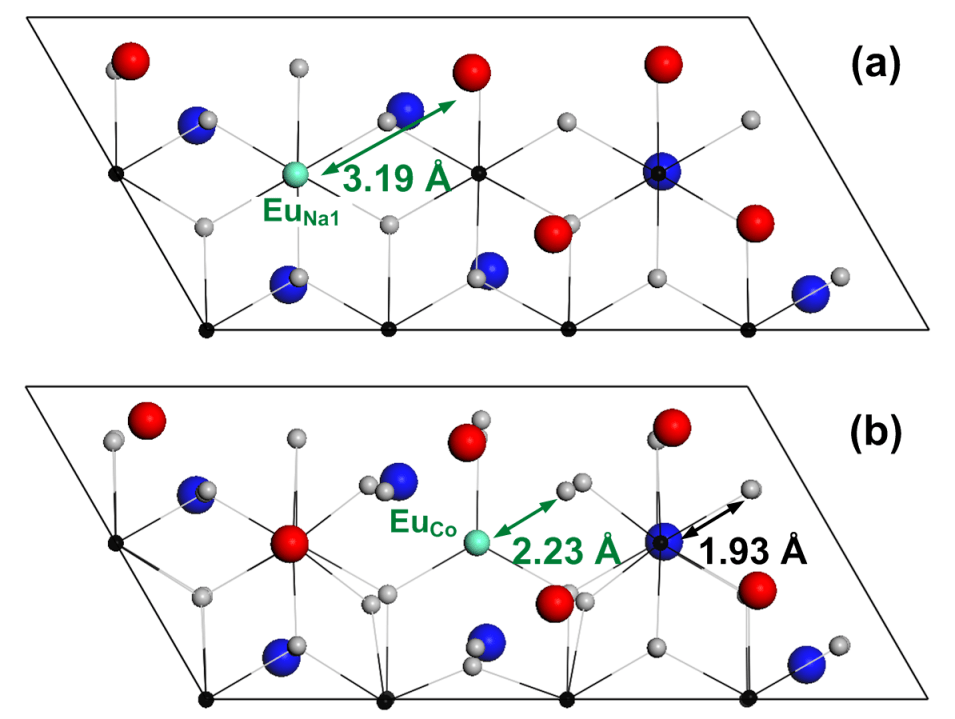}
                \caption{\label{fig:6} Schematic representation of Eu doped \ch{Na_{0.75}CoO2}. Parts (a) and (b) demonstrate the ionic arrangement of \ch{Na_{0.75}CoO2} when doped with  \ch{Eu_{Na 1}^{$\cdot\cdot\cdot$}} and \ch{Eu_{Co}^{$\cdot\cdot\cdot$}} respectively.  \ch{Eu_{Na 1}^{$\cdot\cdot\cdot$}} and \ch{Eu_{Co}^{$\cdot\cdot\cdot$}} are the most and least stable Zn configurations when $E_{\mbox{\scriptsize{Fermi}}}$ = VBM as the case of p-type \ch{Na_{0.75}CoO2}.}
            \end{figure}
            The formation energy of Eu dopants is presented in Fig. \ref{fig:1}(c).
            When $E_{\mbox{\scriptsize{Fermi}}}$ = VBM, \ch{Eu_{Na 1}^{$\cdot\cdot\cdot$}}, the most stable configurations has $E^f$ equal to $-2.64$ eV while \ch{Eu_{Na 2}^{$\cdot\cdot\cdot$}},  \ch{Eu_{Int}^{$\cdot\cdot\cdot$}} and \ch{Eu_{Co}^{$\cdot\cdot\cdot$}} have successively higher $E^f$ of $-2.17$ eV, 0.45 eV and 1.24 eV respectively.
            As $E_{\mbox{\scriptsize{Fermi}}}$ moves across the band gap toward higher values, the $E^f$ of  remains the lowest among all other configurations even after a (3+/2+) and subsequent charge transition at $E_{\mbox{\scriptsize{Fermi}}}$ = 1.53 eV.
            In a short range of $E_{\mbox{\scriptsize{Fermi}}}$ between 1.23 eV and 1.39 eV, \ch{Eu_{Co}}, through rapid transitions, becomes an acceptor and its $E^f$ declines for higher values of $E_{\mbox{\scriptsize{Fermi}}}$.
            However, this process does not make \ch{Eu_{Co}} a stable configuration over other configurations.
            As a result, \ch{Eu_{Na 1}^{$\cdot\cdot\cdot$}} remains the most stable configuration for all values of $E_{\mbox{\scriptsize{Fermi}}}$ that were considered.
            Noticeably, \ch{Eu_{Na 1}} and \ch{Eu_{Na 2}} are never stable in the 1+ charge state and their charge state transits from 2+ to 0 directly indicating that \ch{Eu_{Na 1}} and \ch{Eu_{Na 2}}  are ``negative-U" center dopants.
            Furthermore, \ch{Eu_{Int}} has no transition level for $E_{\mbox{\scriptsize{Fermi}}}$ values up to $\sim{2\si{.eV}}$ and is stable exclusively at 3+ charge state. A schematic representation of \ch{Na_{0.75}CoO2:Eu_{Na 1}} and \ch{Na_{0.75}CoO2:Eu_{Co}} crystal geometry as the most and least stable configurations at $E_{\mbox{\scriptsize{Fermi}}}$ = VBM, is illustrated in Fig. \ref{fig:6}(a) and (b) respectively, presenting configurations with the most and least stable Eu configurations when $E_{\mbox{\scriptsize{Fermi}}}$ = VBM.
            In the case of \ch{Eu_{Na 1}} as shown in Fig. \ref{fig:6}(a), the distance of a particular Eu from its nearest Na neighbor is $3.19\si{.\angstrom}$.
            This distance is larger than the shortest \ch{Na}--\ch{Na} distance of $2.84\si{.\angstrom}$. Furthermore, it is also larger than the shortest dopant--\ch{Na} distance in the \ch{Na_{0.75}CoO2:Zn_{Na 1}} and \ch{Na_{0.75}CoO2:Ni_{Na 1}} systems, demonstrating that more positively charged ions can also be accommodated in the Na layer of these systems.
            Since the ratio of Na2/Na1 is identical to the one of the pristine system, Na2 ions re-arrange themselves by slightly departing from O axis in a Na layer to fit the higher repulsive forces that are caused by \ch{Eu_{Na 1}}.
            As a result, \ch{Eu_{Na 1}} is separated from the Na ions by a greater distance.
            Fig. \ref{fig:6}(b) demonstrates that \ch{Eu_{Co}} causes expansion of $15.5\%$ in \ch{Eu_{Co} - O} bond ($2.33\si{.\angstrom}$) in comparison to the original \ch{Co - O} bond ($1.93\si{.\angstrom}$).
            This substantial expansion has been caused by Eu's larger radius of $0.95\si{.\angstrom}$ than the radius of \ch{Co^{3+}}.
            It seems that \ch{Eu_{Na 1}} causes less lattice distortion than \ch{Eu_{Co}} in this system and thus is more stable.

            \begin{figure}
                \centering
                \includegraphics[width=0.9\columnwidth]{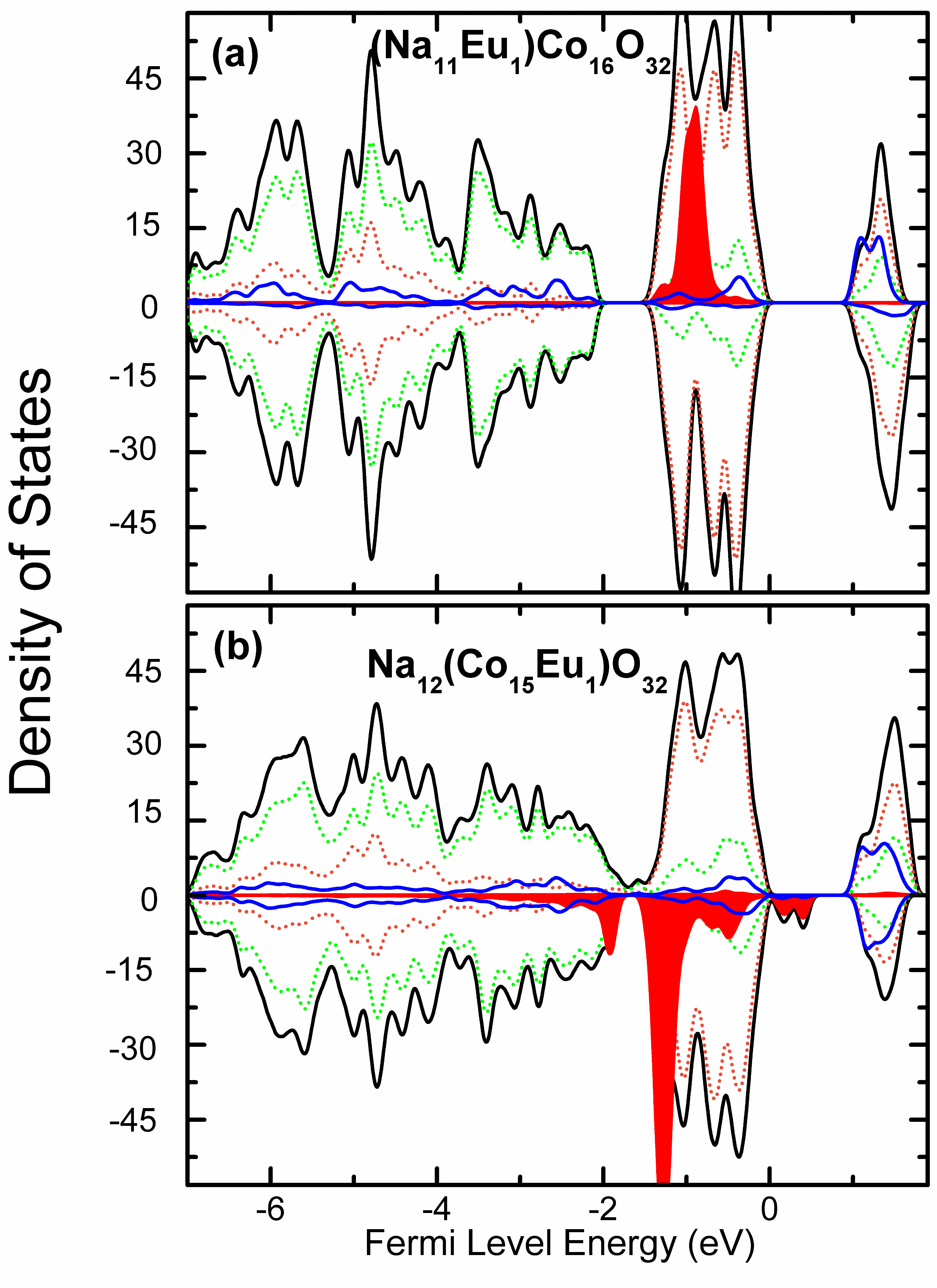}
                \caption{\label{fig:7}Total and partial density of states (PDOS) of  and  doped \ch{Na_{0.75}CoO2} presented in part (a) and (b) respectively. Black, green, orange and blue lines correspond to the total, O $2p$, Co $3d$ and Na $3s$ respectively. The red shaded area represents Eu $4f$ states. For the purpose of clarity, the PDOS of Eu and Na have been magnified.}
            \end{figure}

            Partial density of states of Eu $4f$, Co $3d$, O $2p$ and Na $3s$ electrons along with total DOS are presented in Fig. \ref{fig:7}.
            As in Fig. \ref{fig:7}(a),  $4f$ states are localized in the form of a sharp peak in the middle of the valence band, $\sim 1.5\si{.eV}$ below the Fermi level, hybridizing with O and Co states.
            These states are totally spin polarized and generate a local magnetic moment of $3.45 \mu_B$ on the Eu site. \ch{Eu_{Co}} $4f$ states, as shown in Fig. \ref{fig:7}(b) localize in a major peak at the bottom of the valence band and few smaller peaks in the around Fermi levels.
            The multi-peak nature of the distribution of the \ch{Eu_{Co}} $4f$ states is caused by the stronger crystal field when Eu ions are surrounded by O ions.

        \subsection{Comparison between GGA and GGA+$U$ functionals}
            \begin{figure}
                \centering
                \includegraphics[width=0.9\columnwidth]{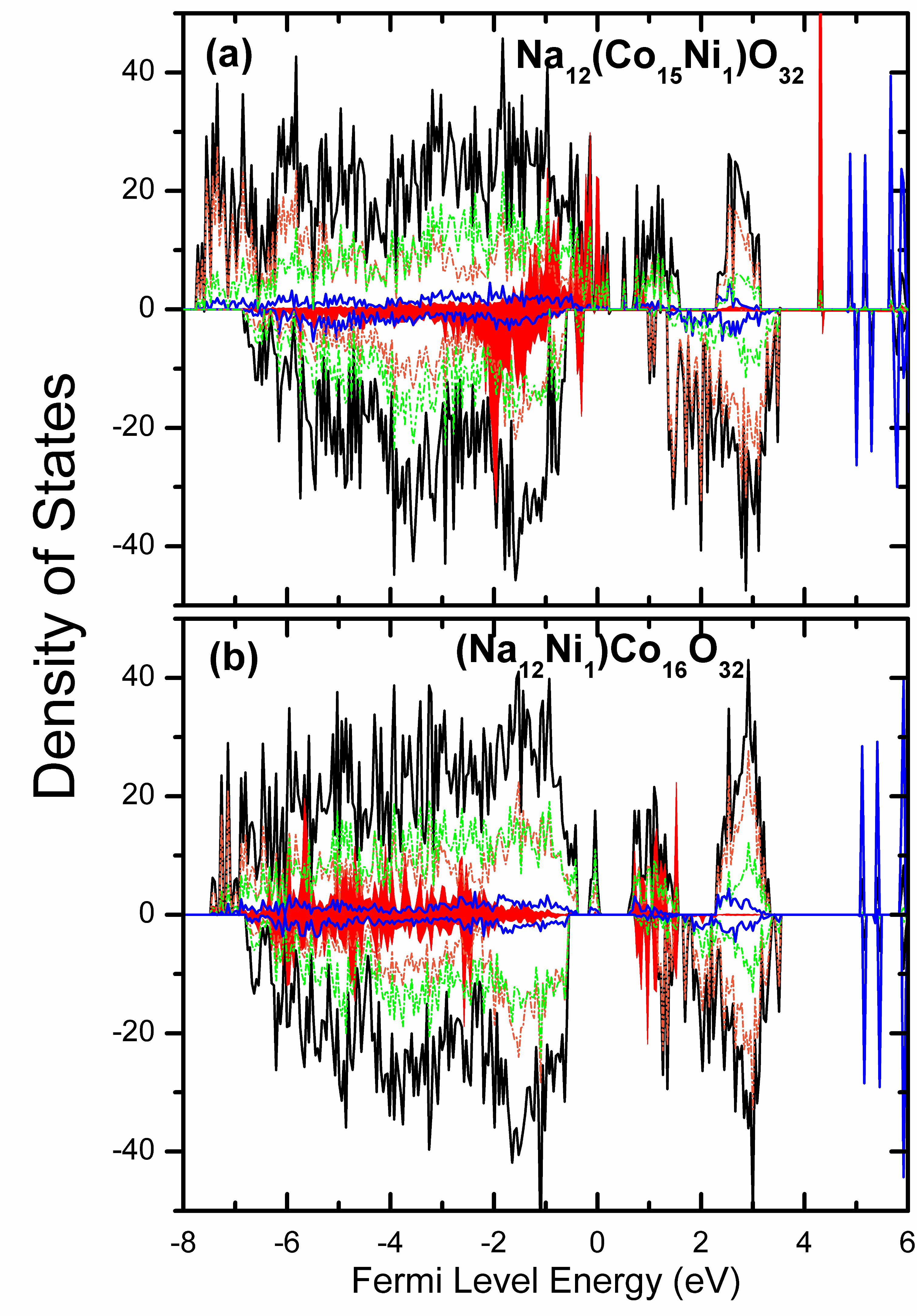}
                \caption{\label{fig:8} Total and partial density of states (PDOS) of  and  doped \ch{Na_{0.75}CoO2} calculated with GGA+$U$ (with effective $U = 4.5\si{.eV}$) presented in part (a) and (b) respectively. Black, green, orange and blue lines correspond to the total, O $2p$, Co $3d$ and Na $3s$ respectively. The red shaded area represents Ni $3d$ states. For the purpose of clarity, the PDOS of Ni and Na have been magnified.}
            \end{figure}
            GGA is well-known for the inaccurate description of the electronic structure of $d$ shell electrons and thus delocalizing them.
            As a solution, introducing an \textit{ad-hoc} on-site orbital dependant Coulomb potential $\left(U\right)$\cite{Anisimov1997} to correct the position of $d$ electrons states is commonly adopted.
            However, introducing a phenomenological parameter such as $U$ renders the approach \textit{non-ab-initio}.
            Here for comparison between the fully \textit{ab-initio} GGA approach and GGA+$U$, the \ch{Na_{0.75}CoO2:Ni} was revisited by introducing a Hubbard term of $U - J = 4.5\si{.eV}$ for the $d$-shell electrons of Co and Ni using VASP code.\cite{Kresse1996, Kresse1999}
            VASP simulation was conducted with the high precision settings for electronic minimization while the thresholds for geometry optimization were fixed to the ones of \ch{DMOL^3} calculations.

            \begin{figure}
                \centering
                \includegraphics[width=0.9\columnwidth]{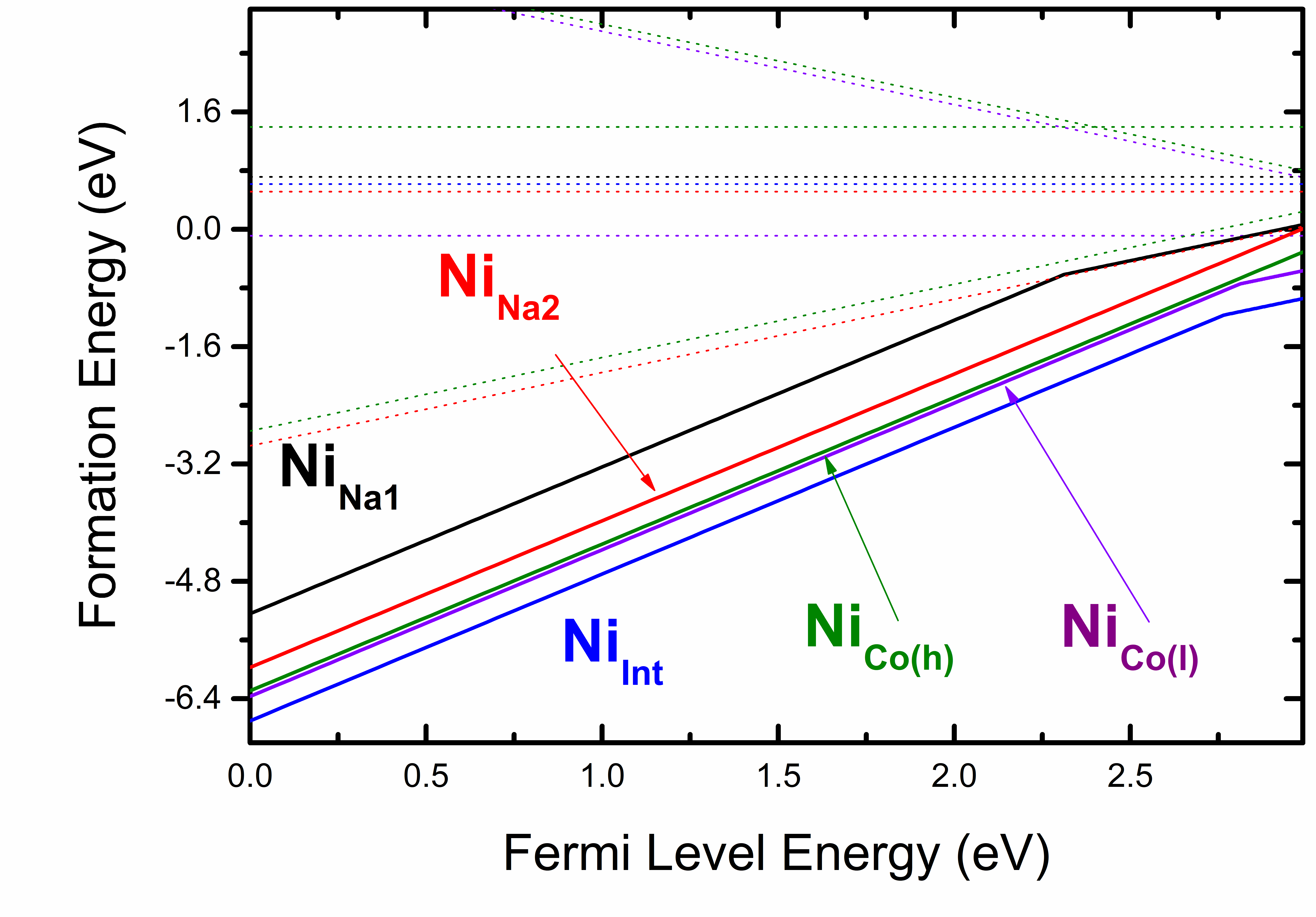}
                \caption{\label{fig:9}The Formation energy ($E^f$) of Ni dopant in \ch{Na_{0.75}CoO2} calculated by GGA+$U$ scheme using VASP code is given. $E^f$ is presented as a function of Fermi level. The dashed lines present negative U charge states that are not stable for any range of $E_{\mbox{\scriptsize{Fermi}}}$.}
            \end{figure}

            Introducing $U$ on $d$-shell electrons results in different charge state and thus the spin numbers among Co ions in pristine \ch{Na_{0.75}CoO2}.
            As a consequence, unlike the results of GGA approach, two different kinds of Co ions are predicted: one with zero spin, and the second type with $\sim 1 e$ spin population.
            This results in Co and Ni $3d$ states, as shown in Fig. \ref{fig:8}, are not degenerate with respect to the spin direction as it was within the GGA approach shown in Fig. \ref{fig:5}.
            As a result, for the Ni doped system five configurations were considered; \ch{Ni_{Na 1}}, \ch{Ni_{Na 2}}, \ch{Ni_{Int}}, \ch{Ni_{Co$\left(h\right)$}} and \ch{Ni_{Co$\left(l\right)$}}, where the last two are Ni ions substituting Co ions with high and low spin numbers.
            The formation energy for these configurations are presented in Fig. \ref{fig:9}.
            Here the lowest $E^f$ belonged to the \ch{Ni_{Int}} configuration with $E^f$ of $-6.70$ eV at $E_{\mbox{\scriptsize{Fermi}}}$ = VBM, followed by the $E^f$ of \ch{Ni_{Co$\left(h\right)$}} and \ch{Ni_{Co$\left(l\right)$}} with $E^f$ of $-6.37$ eV and $-6.29$ eV respectively.
            Within the GGA+$U$ scheme, \ch{Ni_{Na 1}} and \ch{Ni_{Na 2}} configurations were found to be the least stable ones with $E^f$ equal to $-5.36$ eV and $-5.97$ eV respectively.
            The sequence of stabilization of the configurations within the GGA+$U$ scheme differs from the one of the GGA method.
            \ch{Ni_{Int}} which was predicted to be the least stable configuration in GGA method, has become the most stable configuration.
            On similarities between these two approaches, all configurations are stable at Ni's highest oxidation state for $E_{\mbox{\scriptsize{Fermi}}}$ being in the vicinity of VBM.
            \ch{Ni_{Na 1}} is the first configuration to undergo a charge transition and take 1+ charge state.
            The most important insight that can be obtained form GGA+$U$ calculation is that there is only a subtle energy difference between \ch{Ni_{Co$\left(h\right)$}} and \ch{Ni_{Co$\left(l\right)$}} configurations that could not be concluded within the GGA approach.

            To fully reveal the causes of the disagreement between the predictions of GGA and GGA+$U$ approaches, including different stabilization sequence of the configurations, the absolute values of the $E^f$s and the $E_{\mbox{\scriptsize{Fermi}}}$ values for charge state transition, further thorough investigation is required.
            Factors such as the performance of pseudo-potentials and the effect of different values for $U$ on the aforementioned quantities should be the focus of such investigations.

    \section{Discussion}
        The most evident outcome can be seen in the variation of the $E^f$ of different dopants cross a wide range.
        For instance, when $E_{\mbox{\scriptsize{Fermi}}}$ = VBM, \ch{Eu_{Na 1}^{$\cdot\cdot\cdot$}} has a negative $E^f$ of $-2.64$ eV, then followed by \ch{Ni_{Co}^{$\cdot\cdot$}} with $E^f$ of $-2.43$ eV and \ch{Zn_{Co}^{$\cdot\cdot\cdot$}} with $E^f$ of 2.08 eV.
        Since the solubility of a dopant in a host lattice is a direct function of its $E^f$, this trend indicates Eu has a greater solubility in \ch{Na_{0.75}CoO2} than both Ni and Zn.
        It is also well-established that in many oxides, dopant solubility is strongly correlated to the dopant's radius with respect to the ion radii of the host material.\cite{Wilde1998}
        As a result, a higher Eu's solubility may, at first, seem contradictory as it possesses the largest ionic radius among all other studied dopants.
        However, among many other affecting factors, one should note the difference in Na content in the \ch{Na_{0.75}CoO2} doped with \ch{Eu_{Na 1}^{$\cdot\cdot\cdot$}} and the one of \ch{Na_{0.75}CoO2} doped with \ch{Zn_{Co}^{$\cdot\cdot$}} and \ch{Ni_{Co}^{$\cdot\cdot$}} which makes \ch{Eu_{Na 1}^{$\cdot\cdot\cdot$}} more soluble in the host material.
        This has been previously verified by the experimental observation of up to $4\%$ Eu solubility in sodium cobaltate prepared with standard solid state reaction\cite{Tsai2011} with no traces of secondary phases.\cite{Gunawan2007}

        Additionally, it was shown that all dopants tend to be stable at their highest oxidation state.
        The high oxidation state implies an intensive hole-electron recombination and a decrease in carrier concentration, possibly resulting in an increase of resistivity.
        However, one should note that carrier conduction is originated from hopping mechanism in \ch{CoO2} layer and among the studied systems only Eu dopants do not alter the main structural characteristics and chemical composition of this layer.
        As a result, as long as there remain mobile holes in valance band, the metallic conduction should uphold in \ch{Na_{0.75}CoO2:Eu_{Na 1}^{$\cdot\cdot\cdot$}}.
        This argument is supported by experiment where up to $5\%$ rare earth dopants in \ch{Na_{$x$}CoO2}, although increasing resistivity, did not cause metal insulator transition.\cite{Nagira2004}

        In the case of Zn, generally high $E^f$s imply that the Zn solubility should be very limited in \ch{Na_{0.75}CoO2}.
        This point has been experimentally confirmed as attempts to dope more than $1\si{.at}.\%$ Zn in sodium cobaltate lead to the formation of secondary phase.\cite{Tsai2011}
        Since Zn ions have lower oxidation state than Eu ions, at lower doping concentrations, there is less carrier recombination in comparison to the case of Eu dopants and thus higher hole concentration.

        In the case of Ni doped \ch{Na_{0.75}CoO2}, similar to the case of Zn, Ni dopants reside on the \ch{CoO2} layer.
        Having negative formation energy implies that higher concentration of these dopants can be doped in the host material's \ch{CoO2} layer.
        As a result, the chemical and structural characteristics of this layer is greatly affected, hindering the hopping mechanism. This explains the experimental observation of Ni dopant causing metal-insulator transition in \ch{NaCoO2}.\cite{Gayathri2006}
        Similar behavior may also be observed in other transition metal doped \ch{Na_{0.75}CoO2} systems.

        Finally, since Ni and Eu both have partially filled orbitals, Ni and Eu doped \ch{Na_{0.75}CoO2} can be magnetically interesting.
        In the case of Ni as illustrated in Fig. \ref{fig:3}(a) and (b), Ni $3d$ electrons are not magnetically polarized as predicted by GGA scheme.
        This effect is caused by the delocalization of Ni $3d$ electrons and implies that Ni dopants do not significantly alter the magnetic characteristic of the host material.
        However, GGA+$U$ scheme predicts 1.64 e spin population for \ch{Ni_{Int}^{$\cdot\cdot$}} which can be also seen at the asymmetrical \ch{Ni_{Int}} spin distribution of Ni $3d$ states in Fig. \ref{fig:9}(b).
        This make it essentially possible to establish long-range magnetically ordered Ni dopants within the antiferromagnetic network of Co ions.
        Such arrangement results in net magnetization, therefore the opening the possibility of engineering new class of multifunctional \ch{Na_{0.75}CoO2} and warrants further investigation.

        With the Eu being located in Na layer (\ch{Eu_{Na 1}^{$\cdot\cdot\cdot$}}), as shown in Fig. \ref{fig:7}(a), the proximity of $f$ electrons to the ionic nucleus, leaves the $4f$ states highly localized.
        For the \ch{Eu_{Co}^{$\cdot\cdot\cdot$}}, as shown in Fig. \ref{fig:7}(b), the strong crystal field generated by O coordination splits the Eu $4f$ states to bonding and nonbonding states as smaller peaks appear around the major $4f$ peak.
        However, the Eu site still bears significant magnetic moment.
        The preserved magnetic moment of Eu ions in the host materials lattice also opens new horizons for materials engineering with interesting magnetic and thermoelectric effects.

    \section{Conclusion}
        \textit{Ab initio} density functional calculations revealed that the formation energy of the investigated dopants in \ch{Na_{0.75}CoO2} were ranked as following: Eu $<$ Ni $<$ Zn.
        The $E^f$ of these dopants at their most stable configurations ranked as $-2.64$ eV, $-2.35$ eV and 2.08 eV for Eu, Ni and Zn respectively.
        While Eu dopants were highly stabilized when residing in sodium layer, Ni and Zn dopants were more stable when substituting Co in the \ch{CoO2} layer when $E_{\mbox{\scriptsize{Fermi}}}$ = VBM.
        The stability of \ch{Ni_{Co}} and \ch{Zn_{Co}} over their other possible configurations was correlated to the minor lattice distortion caused by Ni and Zn in \ch{CoO2} layer in comparison with the situation when those dopants reside on Na layer.
        Eu $4f$ electrons induced a local magnetic moment of $3.45 \mu_B$.
        The possible implication of the dopant stable oxidation state and crystallographic arrangement on the various properties of sodium cobaltate was discussed and correlated to the thermoelectrical performance of this system.

    \begin{acknowledgements}
        This work was supported by Australian Research Council, Grant Nos.FT100100956 and DP1096185. Computational facility was provided by Australian National Computational Infrastructure and INTERSECT via projects v71 and db1.
    \end{acknowledgements}
    
    \bibliography{Paper2}{}

\begin{thebibliography}{27}%
\makeatletter
\providecommand \@ifxundefined [1]{%
 \@ifx{#1\undefined}
}%
\providecommand \@ifnum [1]{%
 \ifnum #1\expandafter \@firstoftwo
 \else \expandafter \@secondoftwo
 \fi
}%
\providecommand \@ifx [1]{%
 \ifx #1\expandafter \@firstoftwo
 \else \expandafter \@secondoftwo
 \fi
}%
\providecommand \natexlab [1]{#1}%
\providecommand \enquote  [1]{``#1''}%
\providecommand \bibnamefont  [1]{#1}%
\providecommand \bibfnamefont [1]{#1}%
\providecommand \citenamefont [1]{#1}%
\providecommand \href@noop [0]{\@secondoftwo}%
\providecommand \href [0]{\begingroup \@sanitize@url \@href}%
\providecommand \@href[1]{\@@startlink{#1}\@@href}%
\providecommand \@@href[1]{\endgroup#1\@@endlink}%
\providecommand \@sanitize@url [0]{\catcode `\\12\catcode `\$12\catcode
  `\&12\catcode `\#12\catcode `\^12\catcode `\_12\catcode `\%12\relax}%
\providecommand \@@startlink[1]{}%
\providecommand \@@endlink[0]{}%
\providecommand \url  [0]{\begingroup\@sanitize@url \@url }%
\providecommand \@url [1]{\endgroup\@href {#1}{\urlprefix }}%
\providecommand \urlprefix  [0]{URL }%
\providecommand \Eprint [0]{\href }%
\providecommand \doibase [0]{http://dx.doi.org/}%
\providecommand \selectlanguage [0]{\@gobble}%
\providecommand \bibinfo  [0]{\@secondoftwo}%
\providecommand \bibfield  [0]{\@secondoftwo}%
\providecommand \translation [1]{[#1]}%
\providecommand \BibitemOpen [0]{}%
\providecommand \bibitemStop [0]{}%
\providecommand \bibitemNoStop [0]{.\EOS\space}%
\providecommand \EOS [0]{\spacefactor3000\relax}%
\providecommand \BibitemShut  [1]{\csname bibitem#1\endcsname}%
\let\auto@bib@innerbib\@empty
\bibitem [{\citenamefont {Fergus}(2012)}]{Fergus2012}%
  \BibitemOpen
  \bibfield  {author} {\bibinfo {author} {\bibfnamefont {J.~W.}\ \bibnamefont
  {Fergus}},\ }\href {\doibase 10.1016/j.jeurceramsoc.2011.10.007} {\bibfield
  {journal} {\bibinfo  {journal} {J. Eur. Ceram. Soc.}\ }\textbf {\bibinfo
  {volume} {32}},\ \bibinfo {pages} {525} (\bibinfo {year} {2012})}\BibitemShut
  {NoStop}%
\bibitem [{\citenamefont {Koshibae}\ and\ \citenamefont
  {Maekawa}(2001)}]{Koshibae2001}%
  \BibitemOpen
  \bibfield  {author} {\bibinfo {author} {\bibfnamefont {W.}~\bibnamefont
  {Koshibae}}\ and\ \bibinfo {author} {\bibfnamefont {S.}~\bibnamefont
  {Maekawa}},\ }\href {\doibase 236603 10.1103/PhysRevLett.87.236603}
  {\bibfield  {journal} {\bibinfo  {journal} {Phys. Rev. Lett.}\ }\textbf
  {\bibinfo {volume} {87}},\ \bibinfo {pages} {236603} (\bibinfo {year}
  {2001})}\BibitemShut {NoStop}%
\bibitem [{\citenamefont {Terasaki}\ \emph {et~al.}(1997)\citenamefont
  {Terasaki}, \citenamefont {Sasago},\ and\ \citenamefont
  {Uchinokura}}]{Terasaki1997}%
  \BibitemOpen
  \bibfield  {author} {\bibinfo {author} {\bibfnamefont {I.}~\bibnamefont
  {Terasaki}}, \bibinfo {author} {\bibfnamefont {Y.}~\bibnamefont {Sasago}}, \
  and\ \bibinfo {author} {\bibfnamefont {K.}~\bibnamefont {Uchinokura}},\
  }\href {<Go to ISI>://WOS:A1997YH58800001} {\bibfield  {journal} {\bibinfo
  {journal} {Phys. Rev. B}\ }\textbf {\bibinfo {volume} {56}},\ \bibinfo
  {pages} {12685} (\bibinfo {year} {1997})}\BibitemShut {NoStop}%
\bibitem [{\citenamefont {Marianetti}\ and\ \citenamefont
  {Kotliar}(2007)}]{Marianetti2007}%
  \BibitemOpen
  \bibfield  {author} {\bibinfo {author} {\bibfnamefont {C.~A.}\ \bibnamefont
  {Marianetti}}\ and\ \bibinfo {author} {\bibfnamefont {G.}~\bibnamefont
  {Kotliar}},\ }\href {\doibase 176405 10.1103/PhysRevLett.98.176405}
  {\bibfield  {journal} {\bibinfo  {journal} {Phys. Rev. Lett.}\ }\textbf
  {\bibinfo {volume} {98}},\ \bibinfo {pages} {176405} (\bibinfo {year}
  {2007})}\BibitemShut {NoStop}%
\bibitem [{\citenamefont {Foo}\ \emph {et~al.}(2004)\citenamefont {Foo},
  \citenamefont {Wang}, \citenamefont {Watauchi}, \citenamefont {Zandbergen},
  \citenamefont {He}, \citenamefont {Cava},\ and\ \citenamefont
  {Ong}}]{Foo2004}%
  \BibitemOpen
  \bibfield  {author} {\bibinfo {author} {\bibfnamefont {M.~L.}\ \bibnamefont
  {Foo}}, \bibinfo {author} {\bibfnamefont {Y.~Y.}\ \bibnamefont {Wang}},
  \bibinfo {author} {\bibfnamefont {S.}~\bibnamefont {Watauchi}}, \bibinfo
  {author} {\bibfnamefont {H.~W.}\ \bibnamefont {Zandbergen}}, \bibinfo
  {author} {\bibfnamefont {T.}~\bibnamefont {He}}, \bibinfo {author}
  {\bibfnamefont {R.~J.}\ \bibnamefont {Cava}}, \ and\ \bibinfo {author}
  {\bibfnamefont {N.~P.}\ \bibnamefont {Ong}},\ }\href {\doibase 247001
  10.1103/PhysRevLett.92.247001} {\bibfield  {journal} {\bibinfo  {journal}
  {Phys. Rev. Lett.}\ }\textbf {\bibinfo {volume} {92}},\ \bibinfo {pages}
  {247001} (\bibinfo {year} {2004})}\BibitemShut {NoStop}%
\bibitem [{\citenamefont {Mahan}\ and\ \citenamefont {Sofo}(1996)}]{Mahan1996}%
  \BibitemOpen
  \bibfield  {author} {\bibinfo {author} {\bibfnamefont {G.~D.}\ \bibnamefont
  {Mahan}}\ and\ \bibinfo {author} {\bibfnamefont {J.~O.}\ \bibnamefont
  {Sofo}},\ }\href {\doibase 10.1073/pnas.93.15.7436} {\bibfield  {journal}
  {\bibinfo  {journal} {P. Natl. Acad. Sci. USA.}\ }\textbf {\bibinfo {volume}
  {93}},\ \bibinfo {pages} {7436} (\bibinfo {year} {1996})}\BibitemShut
  {NoStop}%
\bibitem [{\citenamefont {Snyder}\ and\ \citenamefont
  {Toberer}(2008)}]{Snyder2008}%
  \BibitemOpen
  \bibfield  {author} {\bibinfo {author} {\bibfnamefont {G.~J.}\ \bibnamefont
  {Snyder}}\ and\ \bibinfo {author} {\bibfnamefont {E.~S.}\ \bibnamefont
  {Toberer}},\ }\href {\doibase 10.1038/nmat2090} {\bibfield  {journal}
  {\bibinfo  {journal} {Nat. Mater.}\ }\textbf {\bibinfo {volume} {7}},\
  \bibinfo {pages} {105} (\bibinfo {year} {2008})}\BibitemShut {NoStop}%
\bibitem [{\citenamefont {Wang}\ \emph {et~al.}(2004)\citenamefont {Wang},
  \citenamefont {Wu}, \citenamefont {Li}, \citenamefont {Chen}, \citenamefont
  {Wang},\ and\ \citenamefont {Luo}}]{Wang2004}%
  \BibitemOpen
  \bibfield  {author} {\bibinfo {author} {\bibfnamefont {N.~L.}\ \bibnamefont
  {Wang}}, \bibinfo {author} {\bibfnamefont {D.}~\bibnamefont {Wu}}, \bibinfo
  {author} {\bibfnamefont {G.}~\bibnamefont {Li}}, \bibinfo {author}
  {\bibfnamefont {X.~H.}\ \bibnamefont {Chen}}, \bibinfo {author}
  {\bibfnamefont {C.~H.}\ \bibnamefont {Wang}}, \ and\ \bibinfo {author}
  {\bibfnamefont {X.~G.}\ \bibnamefont {Luo}},\ }\href {\doibase 147403
  10.1103/PhysRevLett.93.147403} {\bibfield  {journal} {\bibinfo  {journal}
  {Phys. Rev. Lett.}\ }\textbf {\bibinfo {volume} {93}},\ \bibinfo {pages}
  {147403} (\bibinfo {year} {2004})}\BibitemShut {NoStop}%
\bibitem [{\citenamefont {Tsai}\ \emph {et~al.}(2012)\citenamefont {Tsai},
  \citenamefont {Assadi}, \citenamefont {Zhang}, \citenamefont {Ulrich},
  \citenamefont {Tan}, \citenamefont {Donelson},\ and\ \citenamefont
  {Li}}]{Tsai2012}%
  \BibitemOpen
  \bibfield  {author} {\bibinfo {author} {\bibfnamefont {P.~H.}\ \bibnamefont
  {Tsai}}, \bibinfo {author} {\bibfnamefont {M.~H.~N.}\ \bibnamefont {Assadi}},
  \bibinfo {author} {\bibfnamefont {T.~S.}\ \bibnamefont {Zhang}}, \bibinfo
  {author} {\bibfnamefont {C.}~\bibnamefont {Ulrich}}, \bibinfo {author}
  {\bibfnamefont {T.~T.}\ \bibnamefont {Tan}}, \bibinfo {author} {\bibfnamefont
  {R.}~\bibnamefont {Donelson}}, \ and\ \bibinfo {author} {\bibfnamefont
  {S.}~\bibnamefont {Li}},\ }\href {\doibase 10.1021/jp209343v} {\bibfield
  {journal} {\bibinfo  {journal} {J. Phys. Chem. C}\ }\textbf {\bibinfo
  {volume} {116}},\ \bibinfo {pages} {4324} (\bibinfo {year}
  {2012})}\BibitemShut {NoStop}%
\bibitem [{\citenamefont {Nagira}\ \emph {et~al.}(2004)\citenamefont {Nagira},
  \citenamefont {Ito},\ and\ \citenamefont {Hara}}]{Nagira2004}%
  \BibitemOpen
  \bibfield  {author} {\bibinfo {author} {\bibfnamefont {T.}~\bibnamefont
  {Nagira}}, \bibinfo {author} {\bibfnamefont {M.}~\bibnamefont {Ito}}, \ and\
  \bibinfo {author} {\bibfnamefont {S.}~\bibnamefont {Hara}},\ }\href@noop {}
  {\bibfield  {journal} {\bibinfo  {journal} {Mater. Trans.}\ }\textbf
  {\bibinfo {volume} {45}},\ \bibinfo {pages} {1339} (\bibinfo {year}
  {2004})}\BibitemShut {NoStop}%
\bibitem [{\citenamefont {Seetawan}\ \emph {et~al.}(2006)\citenamefont
  {Seetawan}, \citenamefont {Amornkitbamrung}, \citenamefont {Burinprakhon},
  \citenamefont {Maensiri}, \citenamefont {Kurosaki}, \citenamefont {Muta},
  \citenamefont {Uno},\ and\ \citenamefont {Yamanaka}}]{Seetawan2006}%
  \BibitemOpen
  \bibfield  {author} {\bibinfo {author} {\bibfnamefont {T.}~\bibnamefont
  {Seetawan}}, \bibinfo {author} {\bibfnamefont {V.}~\bibnamefont
  {Amornkitbamrung}}, \bibinfo {author} {\bibfnamefont {T.}~\bibnamefont
  {Burinprakhon}}, \bibinfo {author} {\bibfnamefont {S.}~\bibnamefont
  {Maensiri}}, \bibinfo {author} {\bibfnamefont {K.}~\bibnamefont {Kurosaki}},
  \bibinfo {author} {\bibfnamefont {H.}~\bibnamefont {Muta}}, \bibinfo {author}
  {\bibfnamefont {M.}~\bibnamefont {Uno}}, \ and\ \bibinfo {author}
  {\bibfnamefont {S.}~\bibnamefont {Yamanaka}},\ }\href {\doibase
  10.1016/j.jallcom.2005.06.032} {\bibfield  {journal} {\bibinfo  {journal} {J.
  Alloy. Compd.}\ }\textbf {\bibinfo {volume} {407}},\ \bibinfo {pages} {314}
  (\bibinfo {year} {2006})}\BibitemShut {NoStop}%
\bibitem [{\citenamefont {Zhang}\ \emph {et~al.}(2005)\citenamefont {Zhang},
  \citenamefont {Zhao}, \citenamefont {Guo}, \citenamefont {Qiao},
  \citenamefont {Cui}, \citenamefont {Luo}, \citenamefont {Zhang},
  \citenamefont {Yu}, \citenamefont {Shi}, \citenamefont {Zhang}, \citenamefont
  {Zhao},\ and\ \citenamefont {Li}}]{Zhang2005}%
  \BibitemOpen
  \bibfield  {author} {\bibinfo {author} {\bibfnamefont {W.~Y.}\ \bibnamefont
  {Zhang}}, \bibinfo {author} {\bibfnamefont {Y.~G.}\ \bibnamefont {Zhao}},
  \bibinfo {author} {\bibfnamefont {Z.~P.}\ \bibnamefont {Guo}}, \bibinfo
  {author} {\bibfnamefont {P.~T.}\ \bibnamefont {Qiao}}, \bibinfo {author}
  {\bibfnamefont {L.}~\bibnamefont {Cui}}, \bibinfo {author} {\bibfnamefont
  {L.~B.}\ \bibnamefont {Luo}}, \bibinfo {author} {\bibfnamefont {X.~P.}\
  \bibnamefont {Zhang}}, \bibinfo {author} {\bibfnamefont {H.~C.}\ \bibnamefont
  {Yu}}, \bibinfo {author} {\bibfnamefont {Y.~G.}\ \bibnamefont {Shi}},
  \bibinfo {author} {\bibfnamefont {S.~Y.}\ \bibnamefont {Zhang}}, \bibinfo
  {author} {\bibfnamefont {T.~Y.}\ \bibnamefont {Zhao}}, \ and\ \bibinfo
  {author} {\bibfnamefont {J.~Q.}\ \bibnamefont {Li}},\ }\href {\doibase
  10.1016/j.ssc.2005.05.047} {\bibfield  {journal} {\bibinfo  {journal} {Solid
  State Commun.}\ }\textbf {\bibinfo {volume} {135}},\ \bibinfo {pages} {480}
  (\bibinfo {year} {2005})}\BibitemShut {NoStop}%
\bibitem [{\citenamefont {Gayathri}\ \emph {et~al.}(2006)\citenamefont
  {Gayathri}, \citenamefont {Bharathi}, \citenamefont {Sastry}, \citenamefont
  {Sundar},\ and\ \citenamefont {Hariharan}}]{Gayathri2006}%
  \BibitemOpen
  \bibfield  {author} {\bibinfo {author} {\bibfnamefont {N.}~\bibnamefont
  {Gayathri}}, \bibinfo {author} {\bibfnamefont {A.}~\bibnamefont {Bharathi}},
  \bibinfo {author} {\bibfnamefont {V.~S.}\ \bibnamefont {Sastry}}, \bibinfo
  {author} {\bibfnamefont {C.~S.}\ \bibnamefont {Sundar}}, \ and\ \bibinfo
  {author} {\bibfnamefont {Y.}~\bibnamefont {Hariharan}},\ }\href {\doibase
  10.1016/j.ssc.2006.04.030} {\bibfield  {journal} {\bibinfo  {journal} {Solid
  State Commun.}\ }\textbf {\bibinfo {volume} {138}},\ \bibinfo {pages} {489}
  (\bibinfo {year} {2006})}\BibitemShut {NoStop}%
\bibitem [{\citenamefont {Zhang}\ \emph {et~al.}(2006)\citenamefont {Zhang},
  \citenamefont {Zhang}, \citenamefont {Xu}, \citenamefont {Jing},
  \citenamefont {Cao},\ and\ \citenamefont {Zhao}}]{Zhang2006}%
  \BibitemOpen
  \bibfield  {author} {\bibinfo {author} {\bibfnamefont {Z.~Q.}\ \bibnamefont
  {Zhang}}, \bibinfo {author} {\bibfnamefont {J.~C.}\ \bibnamefont {Zhang}},
  \bibinfo {author} {\bibfnamefont {Y.}~\bibnamefont {Xu}}, \bibinfo {author}
  {\bibfnamefont {C.}~\bibnamefont {Jing}}, \bibinfo {author} {\bibfnamefont
  {S.~X.}\ \bibnamefont {Cao}}, \ and\ \bibinfo {author} {\bibfnamefont
  {Y.~G.}\ \bibnamefont {Zhao}},\ }\href {\doibase 045108
  10.1103/PhysRevB.74.045108} {\bibfield  {journal} {\bibinfo  {journal} {Phys.
  Rev. B}\ }\textbf {\bibinfo {volume} {74}},\ \bibinfo {pages} {045108}
  (\bibinfo {year} {2006})}\BibitemShut {NoStop}%
\bibitem [{\citenamefont {Dutta}\ \emph {et~al.}(2007)\citenamefont {Dutta},
  \citenamefont {Battogtokh}, \citenamefont {McKewon}, \citenamefont
  {Vidensky}, \citenamefont {Dutta},\ and\ \citenamefont {Pegg}}]{Dutta2007}%
  \BibitemOpen
  \bibfield  {author} {\bibinfo {author} {\bibfnamefont {B.}~\bibnamefont
  {Dutta}}, \bibinfo {author} {\bibfnamefont {J.}~\bibnamefont {Battogtokh}},
  \bibinfo {author} {\bibfnamefont {D.}~\bibnamefont {McKewon}}, \bibinfo
  {author} {\bibfnamefont {I.}~\bibnamefont {Vidensky}}, \bibinfo {author}
  {\bibfnamefont {N.}~\bibnamefont {Dutta}}, \ and\ \bibinfo {author}
  {\bibfnamefont {I.~L.}\ \bibnamefont {Pegg}},\ }\href {\doibase
  10.1007/s11664-007-0158-9} {\bibfield  {journal} {\bibinfo  {journal} {J.
  Electron. Mater.}\ }\textbf {\bibinfo {volume} {36}},\ \bibinfo {pages} {746}
  (\bibinfo {year} {2007})}\BibitemShut {NoStop}%
\bibitem [{\citenamefont {Delley}(1990)}]{Delley1990}%
  \BibitemOpen
  \bibfield  {author} {\bibinfo {author} {\bibfnamefont {B.}~\bibnamefont
  {Delley}},\ }\href {\doibase 10.1063/1.458452} {\bibfield  {journal}
  {\bibinfo  {journal} {J. Chem. Phys.}\ }\textbf {\bibinfo {volume} {92}},\
  \bibinfo {pages} {508} (\bibinfo {year} {1990})}\BibitemShut {NoStop}%
\bibitem [{\citenamefont {Delley}(2000)}]{Delley2000}%
  \BibitemOpen
  \bibfield  {author} {\bibinfo {author} {\bibfnamefont {B.}~\bibnamefont
  {Delley}},\ }\href {\doibase Pii [s0021-9606(00)30342-7] 10.1063/1.1316015}
  {\bibfield  {journal} {\bibinfo  {journal} {J. Chem. Phys.}\ }\textbf
  {\bibinfo {volume} {113}},\ \bibinfo {pages} {7756} (\bibinfo {year}
  {2000})}\BibitemShut {NoStop}%
\bibitem [{\citenamefont {Perdew}\ \emph {et~al.}(1996)\citenamefont {Perdew},
  \citenamefont {Burke},\ and\ \citenamefont {Wang}}]{Perdew1996}%
  \BibitemOpen
  \bibfield  {author} {\bibinfo {author} {\bibfnamefont {J.~P.}\ \bibnamefont
  {Perdew}}, \bibinfo {author} {\bibfnamefont {K.}~\bibnamefont {Burke}}, \
  and\ \bibinfo {author} {\bibfnamefont {Y.}~\bibnamefont {Wang}},\ }\href
  {\doibase 10.1103/PhysRevB.54.16533} {\bibfield  {journal} {\bibinfo
  {journal} {Phys. Rev. B}\ }\textbf {\bibinfo {volume} {54}},\ \bibinfo
  {pages} {16533} (\bibinfo {year} {1996})}\BibitemShut {NoStop}%
\bibitem [{\citenamefont {Chen}\ \emph {et~al.}(2004)\citenamefont {Chen},
  \citenamefont {Chen}, \citenamefont {Maljuk}, \citenamefont {Kulakov},
  \citenamefont {Zhang}, \citenamefont {Lemmens},\ and\ \citenamefont
  {Lin}}]{Chen2004}%
  \BibitemOpen
  \bibfield  {author} {\bibinfo {author} {\bibfnamefont {D.~P.}\ \bibnamefont
  {Chen}}, \bibinfo {author} {\bibfnamefont {H.~C.}\ \bibnamefont {Chen}},
  \bibinfo {author} {\bibfnamefont {A.}~\bibnamefont {Maljuk}}, \bibinfo
  {author} {\bibfnamefont {A.}~\bibnamefont {Kulakov}}, \bibinfo {author}
  {\bibfnamefont {H.}~\bibnamefont {Zhang}}, \bibinfo {author} {\bibfnamefont
  {P.}~\bibnamefont {Lemmens}}, \ and\ \bibinfo {author} {\bibfnamefont
  {C.~T.}\ \bibnamefont {Lin}},\ }\href
  {http://link.aps.org/doi/10.1103/PhysRevB.70.024506} {\bibfield  {journal}
  {\bibinfo  {journal} {Phys. Rev. B}\ }\textbf {\bibinfo {volume} {70}},\
  \bibinfo {pages} {024506} (\bibinfo {year} {2004})}\BibitemShut {NoStop}%
\bibitem [{\citenamefont {Meng}\ \emph {et~al.}(2008)\citenamefont {Meng},
  \citenamefont {Hinuma},\ and\ \citenamefont {Ceder}}]{Meng2008}%
  \BibitemOpen
  \bibfield  {author} {\bibinfo {author} {\bibfnamefont {Y.~S.}\ \bibnamefont
  {Meng}}, \bibinfo {author} {\bibfnamefont {Y.~Y.}\ \bibnamefont {Hinuma}}, \
  and\ \bibinfo {author} {\bibfnamefont {G.}~\bibnamefont {Ceder}},\ }\href
  {\doibase 104708 10.1063/1.2839292} {\bibfield  {journal} {\bibinfo
  {journal} {J. Chem. Phys.}\ }\textbf {\bibinfo {volume} {128}},\ \bibinfo
  {pages} {104708} (\bibinfo {year} {2008})}\BibitemShut {NoStop}%
\bibitem [{\citenamefont {Van~de Walle}\ and\ \citenamefont
  {Neugebauer}(2004)}]{vandeWalle2004}%
  \BibitemOpen
  \bibfield  {author} {\bibinfo {author} {\bibfnamefont {C.~G.}\ \bibnamefont
  {Van~de Walle}}\ and\ \bibinfo {author} {\bibfnamefont {J.}~\bibnamefont
  {Neugebauer}},\ }\href {\doibase 10.1063/1.1682673} {\bibfield  {journal}
  {\bibinfo  {journal} {J. Appl. Phys.}\ }\textbf {\bibinfo {volume} {95}},\
  \bibinfo {pages} {3851} (\bibinfo {year} {2004})}\BibitemShut {NoStop}%
\bibitem [{\citenamefont {Anisimov}\ \emph {et~al.}(1997)\citenamefont
  {Anisimov}, \citenamefont {Aryasetiawan},\ and\ \citenamefont
  {Lichtenstein}}]{Anisimov1997}%
  \BibitemOpen
  \bibfield  {author} {\bibinfo {author} {\bibfnamefont {V.~I.}\ \bibnamefont
  {Anisimov}}, \bibinfo {author} {\bibfnamefont {F.}~\bibnamefont
  {Aryasetiawan}}, \ and\ \bibinfo {author} {\bibfnamefont {A.~I.}\
  \bibnamefont {Lichtenstein}},\ }\href {\doibase 10.1088/0953-8984/9/4/002}
  {\bibfield  {journal} {\bibinfo  {journal} {J. Phys.: Condens. Matter}\
  }\textbf {\bibinfo {volume} {9}},\ \bibinfo {pages} {767} (\bibinfo {year}
  {1997})}\BibitemShut {NoStop}%
\bibitem [{\citenamefont {Kresse}\ and\ \citenamefont
  {Furthmuller}(1996)}]{Kresse1996}%
  \BibitemOpen
  \bibfield  {author} {\bibinfo {author} {\bibfnamefont {G.}~\bibnamefont
  {Kresse}}\ and\ \bibinfo {author} {\bibfnamefont {J.}~\bibnamefont
  {Furthmuller}},\ }\href {\doibase 10.1103/PhysRevB.54.11169} {\bibfield
  {journal} {\bibinfo  {journal} {Phys. Rev. B}\ }\textbf {\bibinfo {volume}
  {54}},\ \bibinfo {pages} {11169} (\bibinfo {year} {1996})}\BibitemShut
  {NoStop}%
\bibitem [{\citenamefont {Kresse}\ and\ \citenamefont
  {Joubert}(1999)}]{Kresse1999}%
  \BibitemOpen
  \bibfield  {author} {\bibinfo {author} {\bibfnamefont {G.}~\bibnamefont
  {Kresse}}\ and\ \bibinfo {author} {\bibfnamefont {D.}~\bibnamefont
  {Joubert}},\ }\href {\doibase 10.1103/PhysRevB.59.1758} {\bibfield  {journal}
  {\bibinfo  {journal} {Phys. Rev. B}\ }\textbf {\bibinfo {volume} {59}},\
  \bibinfo {pages} {1758} (\bibinfo {year} {1999})}\BibitemShut {NoStop}%
\bibitem [{\citenamefont {Wilde}\ and\ \citenamefont
  {Catlow}(1998)}]{Wilde1998}%
  \BibitemOpen
  \bibfield  {author} {\bibinfo {author} {\bibfnamefont {P.~J.}\ \bibnamefont
  {Wilde}}\ and\ \bibinfo {author} {\bibfnamefont {C.~R.~A.}\ \bibnamefont
  {Catlow}},\ }\href {\doibase 10.1016/s0167-2738(98)00190-8} {\bibfield
  {journal} {\bibinfo  {journal} {Solid State Ionics}\ }\textbf {\bibinfo
  {volume} {112}},\ \bibinfo {pages} {173} (\bibinfo {year}
  {1998})}\BibitemShut {NoStop}%
\bibitem [{\citenamefont {Tsai}\ \emph {et~al.}(2011)\citenamefont {Tsai},
  \citenamefont {Zhang}, \citenamefont {Donelson}, \citenamefont {Tan},\ and\
  \citenamefont {Li}}]{Tsai2011}%
  \BibitemOpen
  \bibfield  {author} {\bibinfo {author} {\bibfnamefont {P.~H.}\ \bibnamefont
  {Tsai}}, \bibinfo {author} {\bibfnamefont {T.~S.}\ \bibnamefont {Zhang}},
  \bibinfo {author} {\bibfnamefont {R.}~\bibnamefont {Donelson}}, \bibinfo
  {author} {\bibfnamefont {T.~T.}\ \bibnamefont {Tan}}, \ and\ \bibinfo
  {author} {\bibfnamefont {S.}~\bibnamefont {Li}},\ }\href {\doibase
  10.1016/j.jallcom.2011.02.045} {\bibfield  {journal} {\bibinfo  {journal} {J.
  Alloy. Compd.}\ }\textbf {\bibinfo {volume} {509}},\ \bibinfo {pages} {5183}
  (\bibinfo {year} {2011})}\BibitemShut {NoStop}%
\bibitem [{\citenamefont {Gunawan}(2007)}]{Gunawan2007}%
  \BibitemOpen
  \bibfield  {author} {\bibinfo {author} {\bibfnamefont {B.~K.}\ \bibnamefont
  {Gunawan}},\ }\emph {\bibinfo {title} {The Development of Rare Earth Doped
  $NaCo_2O_4$ Thermoelectric Materials, The University of New South Wales}},\
  \href@noop {} {\bibinfo {type} {Thesis}} (\bibinfo {year} {2007})\BibitemShut
  {NoStop}%
\end{thebibliography}%
\end{document}